\documentstyle[prb,aps,twocolumn,psfig,epsfig,floats]{revtex}
\begin{document}
\twocolumn[\hsize\textwidth\columnwidth\hsize\csname@twocolumnfalse\endcsname
\title{The mesoscopic proximity effect probed through superconducting
tunneling contacts} 
\author{Frank K. Wilhelm$^{1,2}$ and Alexander A. Golubov$^{3}$}

\address{$^1$Institut f\"ur Theoretische Festk\"orperphysik,
Universit\"at Karlsruhe, D-76128 Karlsruhe, Germany. \\
$^2$ Quantum Transport Group, TU Delft,
P.O. Box 5046, 2600 GA Delft, The Netherlands\\
$^3$ Department of Applied Physics, University of Twente,
7500 AE Enschede, The Netherlands}
\maketitle
\begin{abstract}
We investigate the properties of complex mesoscopic
superconducting-normal hybrid devices, Andreev-Interferometers in the
case, where the current is proped through a superconducting tunneling 
contact whereas the proximity effect is generated by a transparent 
SN-interface. We show within the quasiclassical Green's functions technique,
how the fundamental SNIS-element of the such structures can be mapped onto 
an effective S$^\prime$IS-junction, where S$^\prime$ is the proximised
material 
with an effective energy gap $E_g<\Delta$.  The conductance through such 
a sample at $T=0$ vanishes if $V<\Delta+E_g$, whereas at $T>0$ the
conductance shows a peak at $V=\Delta-E_g$. We propose the 
Andreev-Interferometer, where $E_g$ can be tuned by an external
phase $\phi$ and displays maxima at 0 mod $2\pi$ and minima at $\pi$
mod $2\pi$. This leads to peculiar current-phase-relations, which depart
from a
zero-phase maximum or minimum depending on the bias voltage and can
even show intermediate extreme at $V\approx\Delta$. We propose an
experiment to verify our predictions and show, how our results are
consistent with recent, unexplained experimental results.
\end{abstract}
\pacs{}
]

The proximity effect, although already known for many decades (see
e.g. \onlinecite {degennes:64}), has recently attracted new scientific
interest in the context of mesoscopic normal-superconducting hybrid
structures, which are now experimentally acessible due to progress in
nanofabrication and measurement support technology
\cite{suplatt}. Departing from the properties of single junctions
\cite{BTK,Kastalsky} and the nonmonotonic diffusion conductance of
SN-wires \cite{NazStPRL,GWZ,CourtoisJLTP}, the interest turned to the
possibility of tuning the conductance by an external phase \cite
{NazStPRL,PetrPRL95,Volk} or a loop in the normal part \cite
{GWZ,CourtoisPRL95}. On the other hand, if probed through tunneling
contacts \cite{pothier:96} the conductance is controlled by the DOS
and the induced minigap \cite{GolKupr,belzig:96-2}, which can also be
controlled by a phase \cite{Charlat} and hence opens another channel
for phase controlled conductance of a different sign \cite{ATWZ}. If a
system contains more than one superconducting terminal, a supercurrent
can flow \cite{JLTP,CourtoisPRB}, 
which can be controlled and
reversed externally \cite{WSZ}. The situation becomes more difficult
and in particular time-dependent, if nonequilibrium is created by
applying an external voltage parallel to the junction \cite{LTsns}.

This latter situation is substantially simplified, if one of the contacts is
separated from the rest of the structure by a tunneling barrier. In that
case, the voltage- and phase-drop is concentrated at the barrier and the
problem is essentially split into two parts: The time-dependence of the
phase at the contact and the proximity effect, which determines the
superconducting properties at the normal side of the contact, within the
normal metal. In that case, the physics should be basically identical to the
case of an S$^\prime$IS-junction, where the properties of the
``superconductor'' S$^\prime$ are entirely controlled by the proximity
effect, i.e. we expect a gap of size $E_g<\Delta$ where, 
if the junction is long, $d\gg\xi_0$ $E_g\propto E_{{\rm %
Th}}=D/d^2$, the Thouless energy.
Hence, we will expect the known \cite{Likharev} physics of such S$^\prime$%
IS-contacts: The onset of a tunneling current at $V=\Delta+E_g$ at any $T$
plus the appearance of a current peak at $V=\Delta-E_g$ if $T>0$. The origin 
of this peak can be easiest understood within a semiconductor representation
of the two superconductors, see e.g.\cite{Tinkham}.

Such a structure can in principle be manufactured in a controlled manner. To
the best of our knowledge, this has not yet been realized in Andreev
interferometers. Nevertheless, we
are going to discuss the connection to two experiments: Kutchinsky et al. 
\cite{Kutchinsky} studied the conductance in a T-shaped interferometer with
superconducting contacts in a semiconducting systems, where unwanted
barriers at the interfaces are likely to occur. Antonov et al. \cite{ATWZ}
. in turn
studied a sample with normal tunneling contacts, which might eventually be
connected to superconducting pieces.

\begin{figure}[tbp]
\includegraphics[width=75mm]{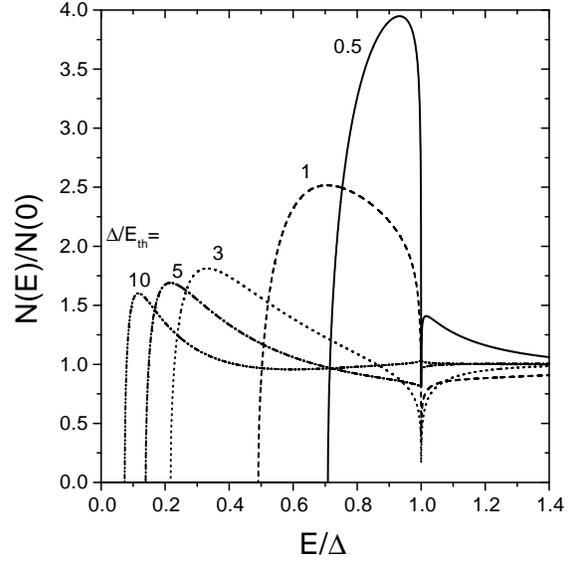}
\caption{DOS in the normal metal at the interface }
\label{fig:dos}
\end{figure}

{\em Model and basic equations.}
Mesoscopic proximity systems are efficiently and quantitatively
described by the quasiclassical Green's functions technique, described
in \onlinecite{review} and its references 4--6, 49, and 50. 
In this approach, the microscopic Gor'kov
equation is reduced to the more handy Usadel equation by various
systematic approximations. At interfaces, this equation is
supplemented by boundary conditions \footnote{In our case, deviations 
from these conditions as discussed in \cite{raimondi} are not likely to occur}
\begin{eqnarray}
p_{F1}^2l_1\hat{G}_1\frac d{dx}\hat{G}_1 &=&p_{F2}^2l_2\hat{G}_2\frac d{dx}%
\hat{G}_2\;,  \label{eq:kuprianov_conditions2} \\
l_2\hat{G}_2\frac d{dx}G_2 &=&t\left[ \hat{G}_2,\hat{G}_1\right] \;.
\end{eqnarray}
These conditions guarantee current conservation. We want to apply them to
the case of small transparencies $t\ll 1$. Here, they enforce that the drop
of phase and voltage is concentrated at the insulating layer \cite
{GolKupr,GolRog}. The current can thus be expressed as an effective
tunneling formula 
\begin{eqnarray}
J &=&\hbox{Re}J_p(V,T)\sin \phi +\hbox{Im}J_p(V,T)\cos \phi +\hbox{Im}%
J_q(V,T)  \nonumber \\
\phi  &=&2eVt+\phi _0  \label{renAB}
\end{eqnarray}
for the current through the interface. Here, the quasiparticle tunneling
current amplitude is 
\begin{eqnarray}
\hbox{Im}\left[J_q(V,T)\right] &=&\frac{G_n}{2e}\int dE\hbox{Re}\left[G_N^R(E)\right]\hbox{Re}\left[G_{BCS}^R(E+V)\right]\nonumber\\
&&\left(\tanh\left( \frac{E+eV}{2T}
\right)+\tanh
\left( \frac{E}{2T}\right) \right)
\label{ssqp}
\end{eqnarray}
and $\hbox{Re}G^R$ gives the quasiparticle DOS. This formula is the
microscopic formulation of the usual Josephson tunneling formula \cite
{parks:Josephson}. 
\begin{figure}[tbp]
\begin{center}
\includegraphics[width=0.75\columnwidth]{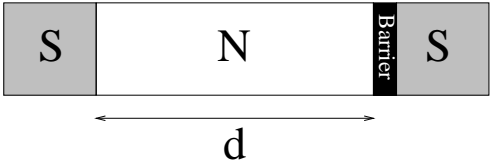}
\end{center}
\caption{Asymmetric SNIS junction
\label{fig:snis}}
\end{figure}
We want to apply this result to the specific case of an SNIS-junction, Fig. 
\ref{fig:snis}. Eq. \ref{renAB} allows to identify this system with an
effective ${\rm S^{\prime }}$ IS-Josephson junction, where the
``superconductor'' ${\rm S^{\prime }}$ is the normal metal layer influenced
by the proximity effect. We can characterize ${\rm S^{\prime }}$ by the
Green's functions at the interface calculated from the Usadel equation
assuming --- in order to be consistent with $R_{{\rm T}}\gg R_{{\rm N}}$ ---
a highly resistive interface and consequently a vanishing phase drop over
the $N$-part. The ``superconductor'' S$^\prime$ has a
gap of size $E_{{\rm G}}\sim \hbox{min}\left(E_{{\rm Th}},\Delta\right)$,
see fig. \ref{fig:dos}. Thus we expect from a semiconductor model that the system shows a
DC supercurrent at $V=0$ and a DC quasiparticle current at $V\ge \Delta +E_{%
{\rm G}}$. Moreover, at finite temperature, a few empty states below $E_{%
{\rm F}}$ and a few quasiparticles above $E_{{\rm F}}$ are available,
enabling transport already at $V\ge \Delta -E_{{\rm G}}$ (see eq. \ref{ssqp}) 
hence leading to a logarithmic quasiparticle
current peak there \cite{Barone}. Unlike the situation in a massive
superconductor, the induced DOS in ${\rm S^{\prime }}$ does not diverge at
the gap edge but has a maximum slightly above $E_G$, see Fig. \ref{fig:dos},
thus we can conclude that also the peak will be smoothened and be slightly
above $\Delta-E_{\rm G}$. 
Additionally, due to BCS singularity in S, another structure
is present in DOS of ${\rm S^{\prime }}$ at $E\sim \Delta $, which is
weakened with increasing thickness $d$ (or decreasing Thouless energy).

{\em Numerical results.}
In order to obtain quantitative results from eqs. \ref{renAB},\ref{ssqp},
the function $\hbox{Im}\left[G^R(d)\right]$ has to be calculated. It is given by the
solution of the Usadel equation 
\[
D\partial^2_x\alpha ^R=2iE\sinh \alpha ^R
\]
with boundary conditions 
\[
\alpha ^R(x=0)=\alpha _S^R=\hbox{Atanh}\left| \frac \Delta E\right| 
\]
at the superconductor and 
\[
\partial _x\left. \alpha ^R\right| _{x=d}=0
\]
at the tunneling barrier, through $G^R(d)=\cosh \alpha ^R$. These nonlinear
equations are in general not solvable analytically. Nevertheless, we find
from a low-energy expansion that $\hbox{Im}\left[\alpha ^R\right]=0$ to all orders, which
indicates the presence of a gap in the spectrum with a sharp edge (at the
convergence radius of the low-energy expansion). At high energies, $E\gg E_{%
{\rm Th}}$, the system is decoupled from the boundary conditions at the
barrier and 
\[
\alpha (d)=4\hbox{Atanh}\left( \tanh (\alpha _S/4)\exp {-\sqrt{-2iE/E_{Th}}}
\right) 
\]
indicating that the deviation from the normal state value is exponentially
cut off at those energies.
This is consistent with our numerical result, Fig. \ref{fig:dos}.

\begin{figure}[tbp]
\includegraphics[width=0.8\columnwidth]{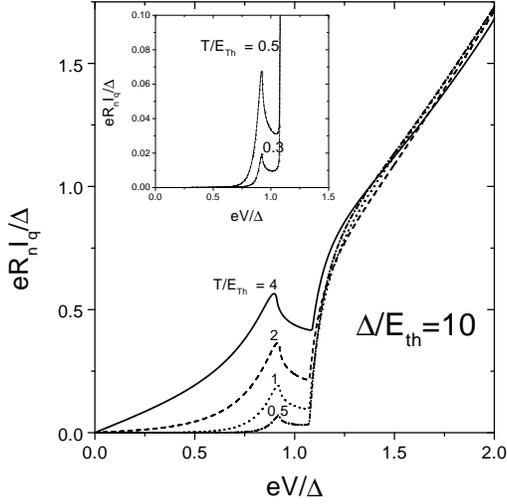}
\caption{I-V-characteristics of an SNIS-junction for $E_{\rm Th}=0.1\Delta$, 
the inset shows the same figure on a
larger scale}
\label{iv5}
\end{figure}

Our qualitative predictions in the preceding section are confirmed by our
numerical results, Fig. \ref{iv5}. As predicted, the peaks grow and smear
out with increasing temperature, but stay visible up to temperatures far
above $E_{{\rm Th}}$. Furthermore, the feature becomes more pronounced if $%
E_G$ is big, i.e. for a shorter junction. 

{\em SNIS Andreev interferometers.}
Even if this type of junction is not prepared on purpose, during
the fabrication process an asymmetric barrier can easily show up
accidentally, e.g. if the N-metal is a highly doped semiconductor and a
Schottky-barrier is likely to occur or if the structure is prepared out of
two layers within a two-step shadow evaporation technique \cite{hweber}.

As a particular example, we discuss a specific set of experiments
\onlinecite{Kutchinsky}. 

\begin{figure}[tbp]
\includegraphics[width=80mm]{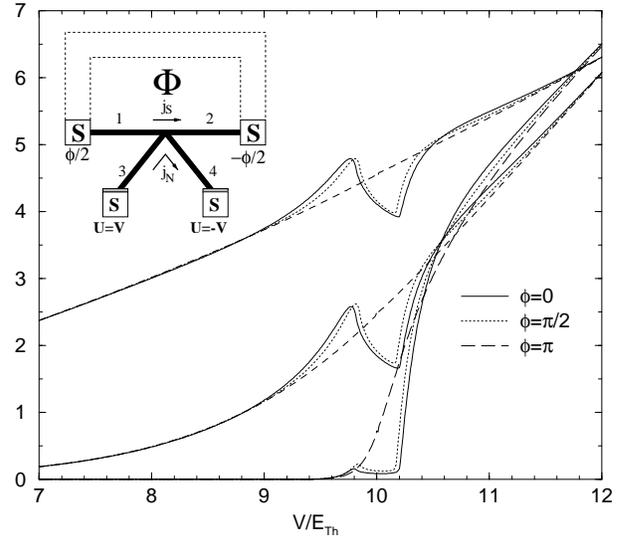}
\caption{I-V-characteristics of the interferometer
shown in the inset for $\Delta =10E_{{\rm Th}}$ at different values of $%
\phi $. Temperatures are (top to bottom) $T=5,1,0.1E_{{\rm Th}}$. All interferometer arms are assumed to be of the same length.}
\label{IVC}
\end{figure}

Unfortunately, 
the interface resistance has not been systematically investigated there, but
a Schottky barrier is likely to occur in this system. 
As a model, we consider the interferometer Fig. \ref{IVC} 
discussed already in \cite{ATWZ} in the case when the tunneling
barriers are strong and all four reservoirs are superconducting.

\begin{figure}[htb]
\epsfig{file=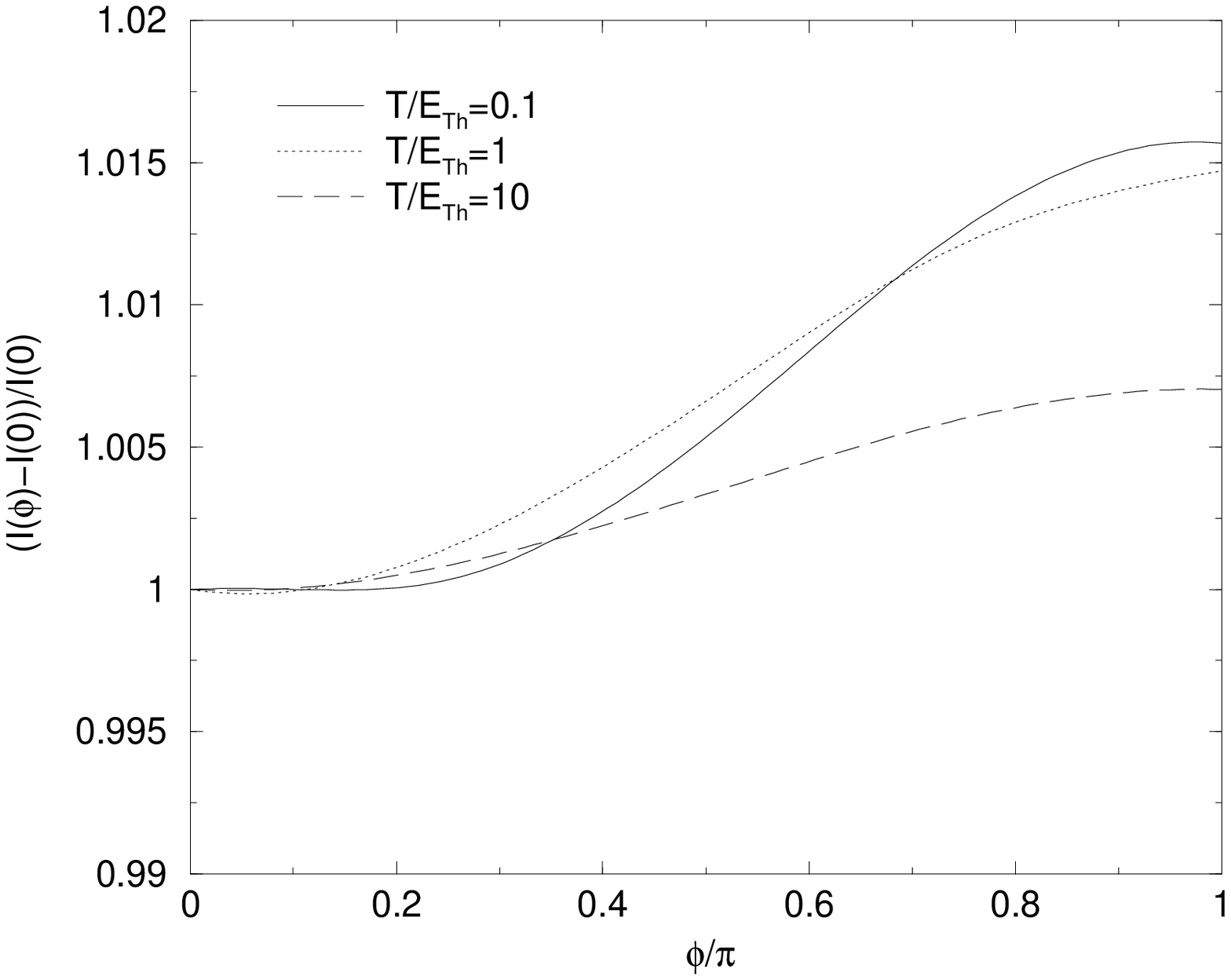,width=0.48\columnwidth}
\hfill
\epsfig{file=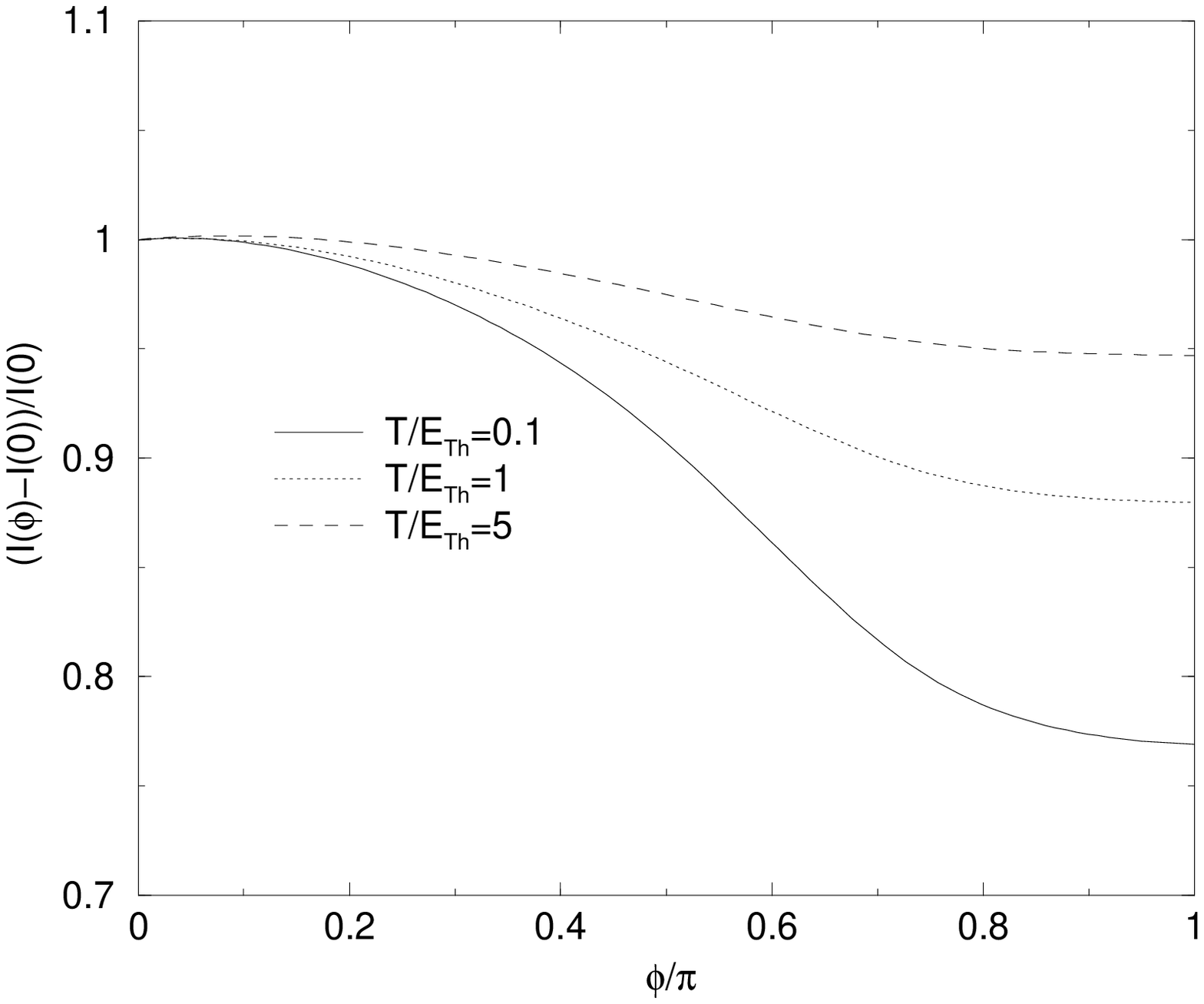,width=0.48\columnwidth}
\epsfig{file=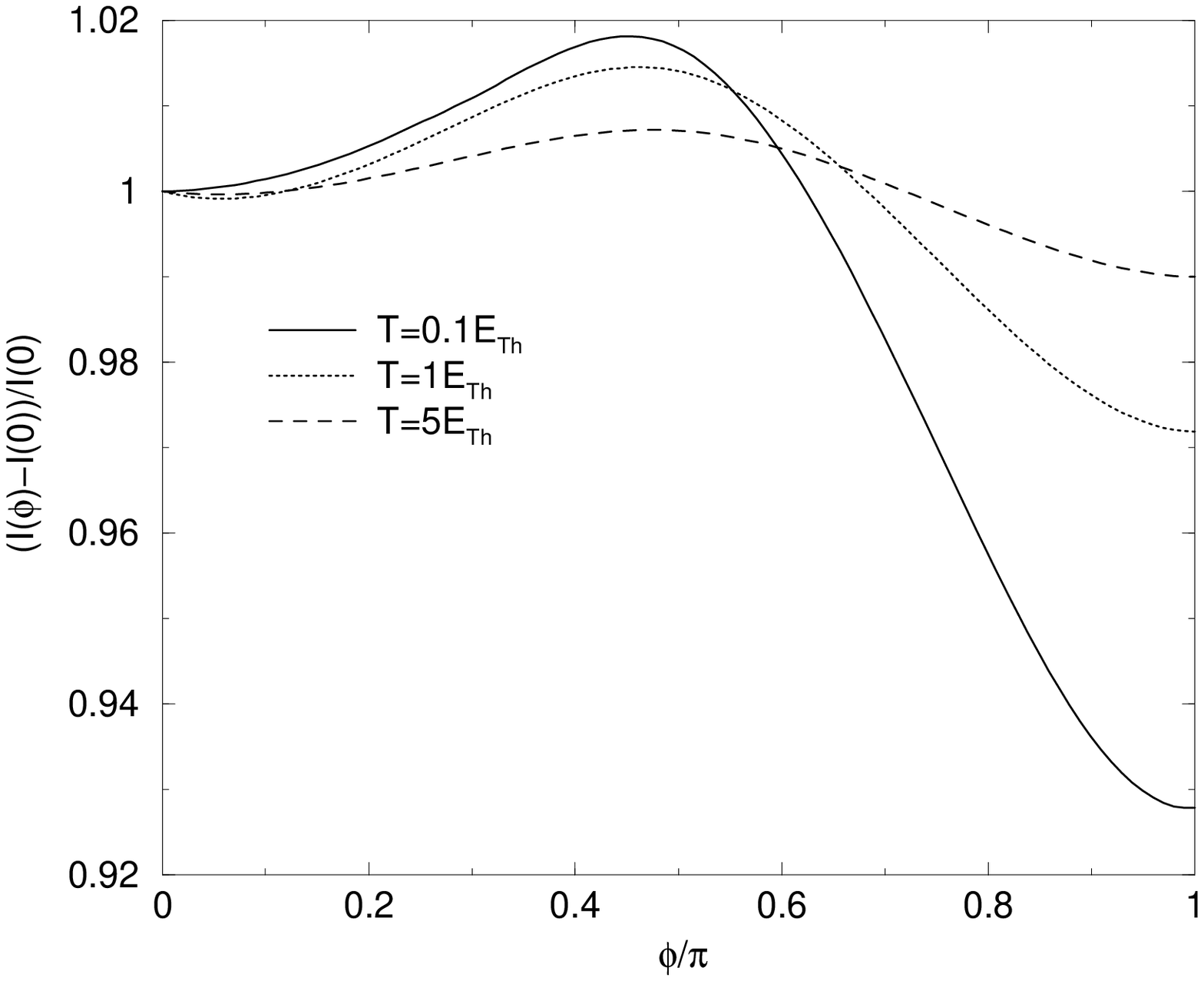,width=0.48\columnwidth}
\hfill
\epsfig{file=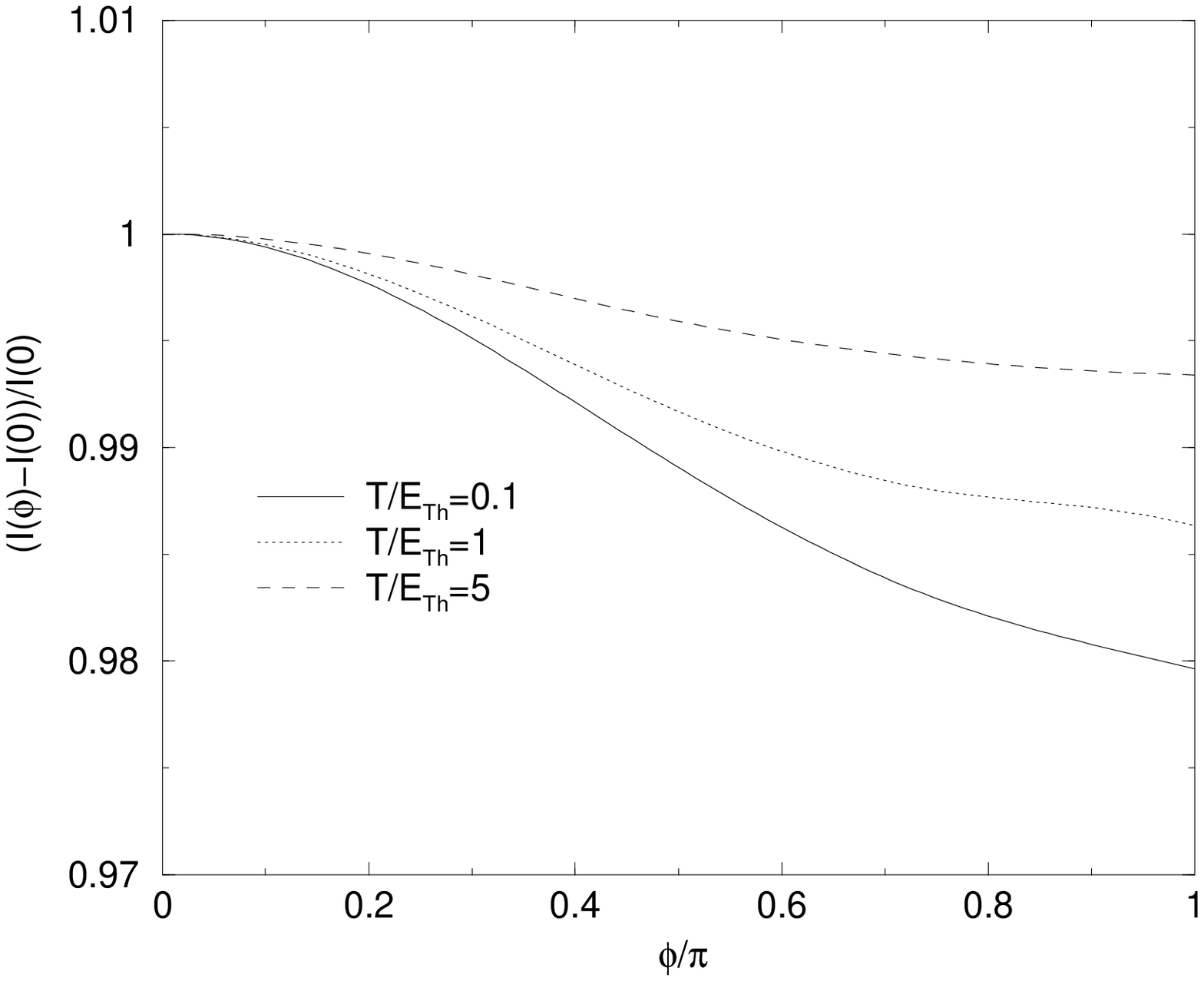,width=0.48\columnwidth}
\caption[Normalized current oscillations probed through a superconducting
contact]{Normalized current oscillations $(I(\phi )-I(0))/I(0)$ at different
temperatures for voltages $V/E_{{\rm Th}}=8.5,9.5,10.5,11.5$}
\label{IVO}
\end{figure}

The phase difference allows to control the strength of the proximity
effect, manifested here in the size of the minigap $E_{\rm G}(\phi)$,
which varies between $E_G^{\rm max}$ at integer and $0$ at
half-integer numbers of flux quanta. The influence of the phase
difference in the interferometer is hence most pronounced for $|\Delta
-E_{\rm G}^{\rm max}|\le V\le |\Delta +E_{\rm G}^{\rm max}|$.  The
I-V characteristics  at a fixed phase, Fig. \ref{IVC}
resembles the form already discussed in Fig. \ref{iv5} but is
slightly smoothened.
At fixed temperatures and voltages, the I-$\phi$ relation 
shows many shapes including zero-field minima and
maxima as well as additional extrema at intermediate phases as depicted
in Fig.\ref{IVO}. This can 
be traced back to the motion of $E_{\rm G}(\phi)$: 
At $V<\Delta -E_G^{{\rm max}}$, a
bigger gap slightly lowers the current (see left upper in fig. \ref{IVO}),
at $\Delta -E_G^{{\rm max}}<V<\Delta $, we are in the vicinity of the
induced peak, which only shows up due to $E_g$, so the current is rather
suppressed by shifting the gap (see right upper in fig. \ref{IVO}). At $\Delta
<V<\Delta +E_G^{{\rm max}}$, the situation is more subtle: The current will
be maximum, if the edge at $\Delta +E_G\approx V$, which will be achieved at
intermediate $\phi $. Due to symmetry reasons, this does not only result
into a phase shift, but into an intermediate maximum. Comparing $\phi =0$
and $\phi =\pi $, one finds that depending on the particular voltage, there
is a competition of the sharpness of the induced gap at $\phi =0$ increasing
on the current above the gap edge but decreasing it below the gap edge,
which have to be traded off and e.g. in Fig. \ref{IVO}, left lower, lead to
a higher current at $\phi =0$. At $V>\Delta +E_g^{{\rm max}}$, both 
peaks in the DOS contribute to the current, which is again 
leads to a zero-phase maximum, right lower.

\begin{figure}[htb]
\includegraphics[width=80mm]{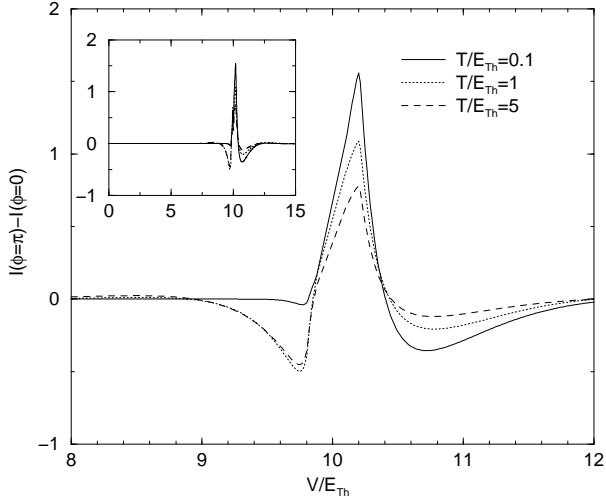}
\caption[Amplitude of the current oscillations]{Amplitude of the ($\phi _0$%
-periodic) current oscillations as a function of the voltage. The inset
shows the full structure}
\label{ampli}
\end{figure}

A similar 
multitude of structures was observed in
the $G(\phi )$ in the interferometer studied in the last section in the
experiments by e.g. by Antonov {\em et al.}, see \cite{ATWZ}. In that paper,
the conductance of an Andreev-interferometer as probed through normal
tunneling contacts was investigated. For technical reasons, small pieces of
Aluminum had to be deposited at the site of the barriers, which may become
superconducting, rendering the structure a superconducting rather than a
normal tunneling contact. As a result, there have been oscillations with
intermediate maxima observed under certain bias conditions, which are
compatible with our predictions \cite{Antonov}. 
The oscillation amplitude, see Fig. \ref{ampli} shows a remarkable peak
structure. In the experiments \cite{Kutchinsky}, this effect will be washed
out due to the 2D-geometry, however, a pronounced splitting of the
conductance peak around $\Delta $ is observed. Remarkably and in agreement
with \onlinecite{Kutchinsky}, the oscillation amplitude in \ref{ampli} only
depends weakly on temperature, although we would have expected a strong
T-dependence at least of the sub-gap peak. This observation in agreement
with the experiments and makes it a likely explanation of the observed peak
splitting. 
Our predictions can be studied in a more genuine setup like 
in the inset of Fig. \ref{IVC}, 
which is also remarkable to another reason: The attached
tunneling contacts cool the distribution function in the normal metal by
removing quasiparticles\cite{Pekola}. This should also influence the
supercurrent between the other two superconducting reservoirs in a way
opposite to \cite{morpurgo}. Whether or not this also leads to $\pi $%
-junction behavior requires more detailed knowledge of the efficiency of the
cooling. The experimental detection of the $\pi$-junction 
along the lines of \onlinecite{Baselmans}
require detailed knowledge of the current-phase-relations 3 and 4
(in that terminology the 
the control line), which is provided by our study.

{\em Summary and Conclusions}
We have discussed the physics of proximity systems probed through a
superconducting tunneling contact. We showed, how these can be understood as
junctions between two different superconductors separated by a tunneling
barrier. This leads to a peculiar current-voltage characteristic containing
a huge step preceded by a small peak at $T>0$. We discussed the
phase-dependence of that current in a typical Andreev-Interferometer and
outlined connections to existing and future experiments.

We would like to acknowledge useful discussions with A.D. Zaikin, G.
Sch\"{o}n, T.M.\ Klapwijk, J.J.A.\ Baselmans, H.\ Weber, T.\ Heikkil\"a, 
O.\ Kuhn, and R. Taboryski. 
This
work was supported by the DFG through SFB 195 and GK 284 and by the EU
through the EU-TMR ``Superconducting Nanocircuits''.


\begin{references}
\bibitem{degennes:64}  P.G.\ de Gennes, Rev.\ Mod.\ Phys.\ {\bf 36}, 225
(1964).

\bibitem{suplatt}  Special issue of {\em Superlattices and Microstructures}, 
{\bf 25}, 627--1288 (1999) ed. P.F.Bagwell, and references therein.

\bibitem{BTK}  G.E.\ Blonder, M.\ Tinkham, and T.M.\ Klapwijk, Phys. Rev. B 
{\bf 25}, 4515 (1982).

\bibitem{Kastalsky}  A. Kastalsky {\em et al.},
Phys. Rev. Lett. {\bf 67}, 3026 (1991).

\bibitem{NazStPRL}  Yu.V.\ Nazarov and T.H.\ Stoof, Phys.\ Rev.\ Lett.\ {\bf %
76}, 823 (1996).

\bibitem{GWZ}  A.A.\ Golubov, F.K.\ Wilhelm, and A.D.\ Zaikin, Phys.\  Rev.\
B {\bf 55}, 1123 (1997).

\bibitem{CourtoisJLTP}  H.\ Courtois {\em et al.}, J Low Temp. Phys. {\bf 116%
}, 187 (1999).

\bibitem{PetrPRL95}  V.T.\ Petrashov {\em et al.}, Phys.\ Rev.\ Lett.\ {\bf %
74}, 5268 (1995).

\bibitem{Volk}  A. F. Volkov, A. F. N. Allsopp, and C. J. Lambert, J. Phys.
Condens. Matter {\bf 8}, L45 (1996).

\bibitem{CourtoisPRL95}  H.\ Courtois {\em et al.}, Phys.\ Rev.\ Lett.\ {\bf %
76}, 130 (1996).

\bibitem{pothier:96}  S.\ Gu\'{e}ron {\em et al.}, Phys.\ Rev.\ Lett.\ {\bf %
77}, 3025 (1996).


\bibitem{GolKupr}  A.A.\ Golubov and M.Yu.\ Kupriyanov, J. Low Temp. Phys. 
{\bf 70}, 83 (1988); JETP {\bf 69}, 805 (1989).

\bibitem{belzig:96-2}  W.\ Belzig, C.\ Bruder, and G.\ Sch\"{o}n, Phys.\
Rev.\ B {\bf 54}, 9443 (1996).

\bibitem{Charlat}  F.\ Zhou {\em et al.}, J.\ Low Temp.\ Phys.\ {\bf 110},
841 (1998).

\bibitem{ATWZ}  V.N.\ Antonov, H.\ Takayanagi, F.K.\ Wilhelm, and A.D.\
Zaikin, submitted.

\bibitem{JLTP}  F.K.\ Wilhelm, A.D.\ Zaikin, and G.\ Sch\"{o}n, J.\ Low
Temp.\ Phys.\ {\bf 106}, 305 (1997).\ 

\bibitem{CourtoisPRB}  H.\ Courtois, Ph.\ Gandit, and B.\ Pannetier, Phys.\
Rev.\ B {\bf 52}, 1162 (1995).

\bibitem{WSZ}  F.K.\ Wilhelm, G.\ Sch\"{o}n, and A.D.\ Zaikin, Phys.\  Rev.\
Lett.\ {\bf 81}, 1682 (1998); to appear in the proceedings of the {\em %
Rencontres de Moriond 1999} conference; to appear in Physica B.

\bibitem{LTsns}  P.\ Dubos {\em et al.}, to appear in Physica B.

\bibitem{Likharev}  K.K.\ Likharev, {\em Dynamics of Josephson Junctions and
Circuits} (Gordon and Breach, N.Y., 1986).

\bibitem{Tinkham} M.\ Tinkham, {\em Introduction to Superconductivity}, 
2nd ed., (McGraw Hill, Singapore, 1996). 

\bibitem{Kutchinsky}  J.\ Kutchinsky {\em et al.}, Phys. Rev. B {\bf 56},
R2932 (1997).

\bibitem{review}  W.\ Belzig, F.K.\ Wilhelm, C.\ Bruder, G.\ Sch\"{o}n, and
A.D.\ Zaikin, Superlattices and Microstructures {\bf 25}, 1251 (1999).

\bibitem{raimondi} C.J.\ Lambert, R.\ Raimondi, V.\ Sweeney, and A.F.\ Volkov,
Phys. Rev. B {\bf 55}, 6015 (1997).


\bibitem{GolRog}  A.A.\ Golubov {\em et al.}, Phys. Rev. B {\bf 51}, 1073
(1995).

\bibitem{parks:Josephson}  B.\ Josephson in R.D.\ Parks, {\em %
Superconductivity}, (Dekker, N.Y., 1969).

\bibitem{Barone}  A. Barone, G. Paterno {\em Physics and applications of the
Josephson effect} (Wiley, N.Y., 1982).

\bibitem{hweber}  H.\ Weber, private communication.

\bibitem{Antonov}  V.N.\ Antonov, private communication.

\bibitem{Pekola}  M.M. Leivo, J.P. Pekola, and D.V. Averin, Appl. Phys.
Lett. {\bf 68}, 1996 (1996).

\bibitem{morpurgo}  A.\ Morpurgo, B.J.\ van Wees, and T.M.\ Klapwijk, Appl.\
Phys.\ Lett.\ {\bf 72}, 966 (1998).

\bibitem{Baselmans}  J.J.A.\ Baselmans, A.F.\ Morpurgo, B.J.\ van Wees, and
T.M.\ Klapwijk, Nature {\={3} 397}, 45 (1999).


\end{references}
\end{document}